# Photonic Analogue of Two-dimensional Topological Insulators and Helical One-Way Edge Transport in Bi-Anisotropic Metamaterials

Alexander B. Khanikaev†, S. Hossein Mousavi, Wang-Kong Tse, Mehdi Kargarian, Allan H. MacDonald, and Gennady Shvets‡

*Department of Physics, The University of Texas at Austin, 2515 Speedway C1500, Austin TX 78712, USA*

†khanikaev@physics.utexas.edu; ‡gena@physics.utexas.edu

Recent progress in understanding the topological properties of condensed matter has led to the discovery of time-reversal invariant topological insulators. Because of limitations imposed by nature, topologically non-trivial electronic order seems to be uncommon except in small-band-gap semiconductors with strong spin-orbit interactions. In this Article we show that artificial electromagnetic structures, known as metamaterials, provide an attractive platform for designing photonic analogues of topological insulators. We demonstrate that a judicious choice of the metamaterial parameters can create photonic phases that support a pair of helical edge states, and that these edge states enable one-way photonic transport that is robust against disorder.

## 1. Introduction

The recent discovery of two [1, 2] and three [3, 4, 5, 6, 7] dimensional topological insulators has stimulated interest in nontrivial topological phases. Although success in identifying topological insulator behavior among strongly spin-orbit coupled electronic systems has been impressive [8, 9, 10, 11], the property is still relatively uncommon. Realizing nontrivial topological phases in other materials systems [12, 13, 14, 15] is therefore highly desirable. Topological phases in fact have been explored in optical systems. In particular, photonic counterparts of quantum Hall edge states have been predicted [13, 14, 16, 17] and experimentally observed [18, 19] in systems with broken time-reversal symmetry. Separately, two-dimensional arrays of coupled resonator optical waveguides (CROW) with time-reversal symmetry were recently proposed [15] as a realization of topologically protected optical delay lines. In this paper, we propose a new approach to topological meta-material design that is based on a precise mapping between a photonic system and an electronic topological-insulator [1, 2]. We show that topologically nontrivial phases can be constructed in periodic metamaterials, referred to below to as photonic meta-crystals, by inducing a coupling between photon polarization states and photon momentum, thereby creating a controllable photonic analog of spin-orbit coupling.

Optical spin-orbit coupling is embodied in the transversality constraints ($\nabla \cdot \mathbf{E} = 0$ and $\nabla \cdot \mathbf{H} = 0$) of the free electromagnetic field. A variety of manifestations of this coupling, including the optical Magnus or Hall effect and the Imbert-Fedorov shift, have been predicted [20, 21, 22, 23] and verified [24, 25]. Especially pronounced topological effects are expected in periodic dielectric media like photonic crystals in which ray optics can be dramatically altered [22, 26, 27]. More generally, enhanced spin-orbit coupling can be achieved in bi-anisotropic materials with magneto-electric coupling [ 28, 29]. When magneto-electric coupling is present electric polarization induced by an incident electromagnetic wave excites microscopic magnetic moments in the material and induces magnetization oscillations at the incident field frequency.

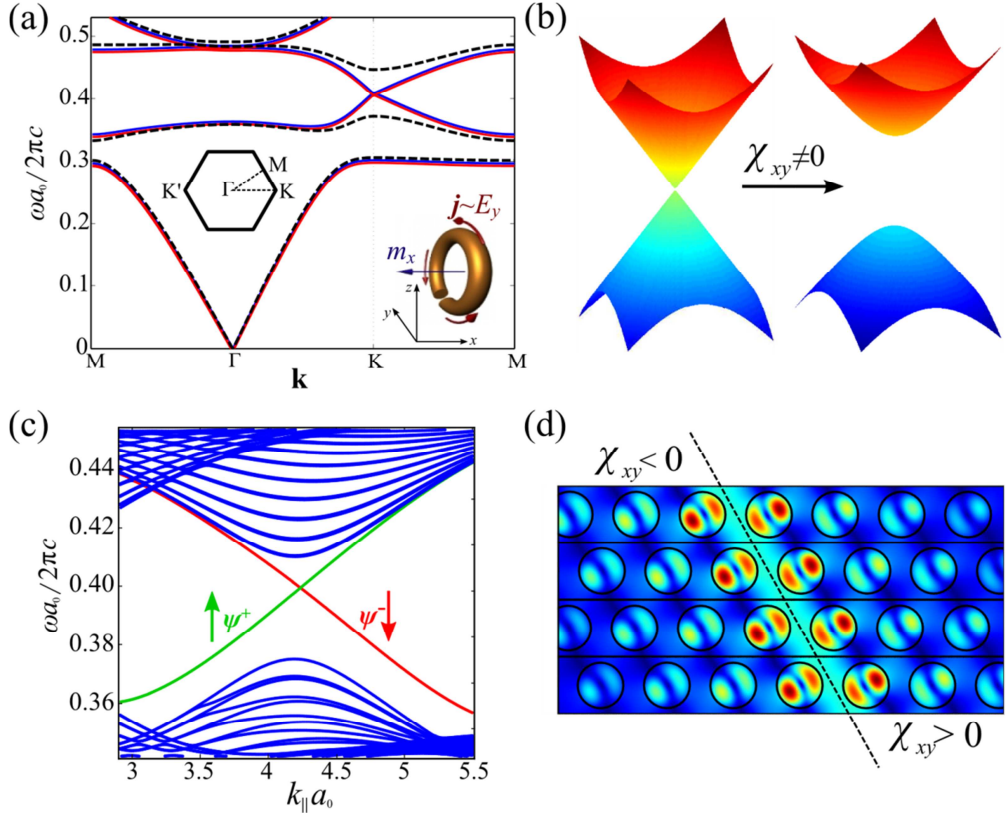

Figure 1: (a) Band structure of the photonic analogue of a topological insulator. The red and blue solid lines show degenerate TE and TM modes in a system in which the magneto-electric coupling tensor $\hat{\chi}$ vanishes. The small spectral shift in the figure is for visualization purposes only. The black dashed lines illustrate removing the Dirac degeneracy by adding bi-anisotropy with $\chi_{xy} = -\chi_{yx} = 0.5$. (No artificial spectral shifts are present in this case.) Note that a complete photonic band gap is present at lower frequency even in the $\chi_{ij}=0$ case. (b) Energy surfaces illustrating degeneracy removal at the K-point of the Brillouin zone for $\chi_{xy} = -\chi_{yx} = 0.1$. (c) Dispersion of the spin-up (green) and spin-down (red) helical edge states supported by a bi-anisotropic domain wall. The blue lines illustrate supercell bulk photonic states. (d) Field distribution $|\psi^{\pm}| = |E_z \pm H_z|$ of spin-up and spin-down edge states supported by a bi-anisotropic domain wall. The difference in the temporal evolution between the spin-up and spin-down modes can be seen from animations of the $\psi^{\pm}$ wavefunctions given in the Multimedia Supplement.

The electromagnetic response of linear bi-anisotropic systems is described macroscopically by the constitutive relations $\boldsymbol{D} = \hat{\epsilon}\boldsymbol{E} + i\hat{\chi}\boldsymbol{H}$ and $\boldsymbol{B} = \hat{\mu}\boldsymbol{H} - i\hat{\chi}^T\boldsymbol{E}$ [ 30, 31]. For real permittivity $\hat{\epsilon}$, permeability $\hat{\mu}$ and bi-anisotropy $\hat{\chi}$ tensors, the form of these constitutive relations ensures time-reversal symmetry and energy conservation [ 30]. To first order in $\hat{\chi}$, these constitutive relations can be rewritten using Maxwell's equations in reciprocal space to obtain a new set of relations, $\boldsymbol{D} = \hat{\epsilon}\boldsymbol{E} + i\hat{\chi}\hat{\mu}^{-1}\boldsymbol{k} \times \boldsymbol{E}/k_0$ and $\boldsymbol{B} = \hat{\mu}\boldsymbol{H} + i\hat{\chi}^T\hat{\epsilon}^{-1}\boldsymbol{k} \times \boldsymbol{H}/k_0$, which explicitly reveal the coupling between photon momentum $\boldsymbol{k}$ and photon polarization states.

One of the most well-known consequences of bi-anisotropy in naturally occurring bi-anisotropic materials is the polarization rotation of linearly polarized light [32]. This property is referred to as optical activity and can be found in materials that contain chiral molecules. Although optically active materials are quite abundant in Nature, their degree of optical activity is generally very weak ($|\chi_{ij}| \ll 1$). However, because of rapid recent advances in the field of metamaterials strong magneto-electric coupling can now be engineered at frequencies from microwave to visible with artificial electromagnetic atoms of rather simple geometry, for example with the use of split-ring resonators [31] (Fig. 1a inset) or plasmonic helices [33]. In this case magneto-electric coupling can be strong ($\chi_{ij} \sim \epsilon_{ij}$ and $\mu_{ij}$) and offer an alternative approach to negative refraction and superlensing [34].

There is a problem in optics that is absent in electronic systems. Even in the absence of bi-anisotropy, the degeneracy between photon polarization states (the photon analogue of spin states) is generally broken because all naturally occurring materials respond differently to the electric and magnetic field components of light ($\epsilon \neq \mu$). In two-dimensional structures, for instance, the difference between electric and magnetic polarizabilities results in different photon-band dispersions for TE and TM modes. Employing a metamaterial structure that is $\epsilon/\mu$-matched throughout, so that the wave impedance everywhere is equal to the free-space value, can cure this problem. Metamaterials with this property have recently attracted significant attention in the context of "transformation optics" [35, 36, 37], and were used to emulate various phenomena of celestial mechanics in optics [38, 39] and to design optical invisibility cloaks [36, 40].

## 2. Photonic topological insulator

As a model for our photonic implementation of topological insulators, we take a two-dimensional periodic metamaterial – a photonic meta-crystal – with a hexagonal lattice of $\epsilon/\mu$-matched rods. The hexagonal symmetry of the structure implies Dirac-cone linear photonic dispersion near the K and K′ corner points of the Brillouin zone [16] (see inset to Fig. 1a) that simplifies mapping to electronic systems [1, 2]. In a general two-dimensional system made of optically passive isotropic media, Maxwell equations split into two independent scalar wave equations corresponding to the transverse magnetic (TM) mode for the electric field $E_z$ and transverse electric (TE) mode for the magnetic field $H_z$. Aiming toward two-dimensional metamaterials with a maximally simplified geometry (see Section 4 below), we consider a meta-crystal with columns made of uniaxial anisotropic metamaterial with the principal axis along the z direction $\hat{\epsilon} = \mathrm{diag}\{\epsilon_\perp, \epsilon_\perp, \epsilon_{zz}\}$ and $\hat{\mu} = \mathrm{diag}\{\mu_\perp, \mu_\perp, \mu_{zz}\}$. To restore the photon polarization degeneracy, we assume matched values of the electric permittivity and magnetic permeability of all the structure's constituents. In what follows we take $\epsilon_\perp = \mu_\perp = 14$ and $\epsilon_{zz} = \mu_{zz} = 1$ for the rods of radius $r_0 = 0.34 a_0$, where $a_0$ is the lattice constant, and $\epsilon_\perp = \mu_\perp = 1$ and $\epsilon_{zz} = \mu_{zz} = 1$ for the background.

A typical band diagram for the optically passive ($\hat{\chi}=0$) crystal obtained by the plane-wave expansion method is shown in Fig. 1a. One can see that the band structure contains two identical overlaid spectra corresponding to the degenerate TE and TM modes, so that any linear combination of these modes is also a solution of the eigenvalue problem. In addition, a pair of degenerate Dirac points corresponding to TE and TM polarizations appears for the second and third doubly degenerate bands crossing at the K point (Figs. 1a and 1b). Note that the band structure around the K and K' points is identical because of time-reversal symmetry of the system.

We now introduce photonic spin-orbit coupling by adding bi-anisotropy. To mimic the effect of gap opening due to strong spin-orbit coupling in topological insulators [1, 2] we consider a particular form of bi-anisotropy that gives rise to mixing between TE and TM modes. The TE-TM degeneracy at the Dirac point will be lifted thus producing a photonic band gap. For our particular metamaterial structure, we find that the only way to induce TE-TM mode coupling that results in the opening of a complete gap at the Dirac points is to introduce an anti-symmetric magneto-electric tensor $\chi_{xy} = -\chi_{yx} \neq 0$ with all other elements $\chi_{ij}=0$ (see Supplement A for details). For this case Maxwell's equations in first order in $\chi_{xy}$ take the form:

$$\left(k_0^2 \mu_{zz} + \nabla_\perp \frac{1}{\epsilon_\perp} \nabla_\perp\right) H_z = \left[\nabla_\perp \left(\frac{-i\chi_{xy}}{\epsilon_\perp \mu_\perp}\right) \times \nabla_\perp\right]_z E_z,$$

$$\left(k_0^2 \epsilon_{zz} + \nabla_\perp \frac{1}{\mu_\perp} \nabla_\perp\right) E_z = \left[\nabla_\perp \left(\frac{-i\chi_{xy}}{\epsilon_\perp \mu_\perp}\right) \times \nabla_\perp\right]_z H_z,$$

which clearly shows the coupling between the TE and TM polarizations. The form of bi-anisotropy used also has the advantage of being experimentally simple to implement (See section 4). The black dashed lines in Fig. 1a show the resulting band structure of the photonic meta-crystal. (The method used to solve the wave equation in the presence of this form of bi-anisotropy is detailed in Supplement A.) In Fig. 1b, we illustrate the band gap opening at the Dirac point when the magneto-electric coupling $\chi_{xy}$ is turned on.

At this point it is instructive to construct an effective "low-energy" Hamiltonian for the system and show how it maps to the electronic Hamiltonian of topological insulators. This can be done by truncating the plane-wave basis to six plane-waves corresponding to the three reciprocal lattice vectors closest to K and K' points and then using the photonic $\boldsymbol{k} \cdot \boldsymbol{p}$ approximation [41, 16]. This procedure results in the following effective Hamiltonian valid in the vicinity of the K and K' points (see Supplement A)

$$\widehat{H} = \begin{bmatrix} v_D(\hat{t}_z\hat{\sigma}_x k_x + \hat{\sigma}_y k_y) & \zeta \hat{t}_z \hat{\sigma}_z \\ \zeta \hat{t}_z \hat{\sigma}_z & v_D(\hat{t}_z\hat{\sigma}_x k_x + \hat{\sigma}_y k_y) \end{bmatrix}. \quad (1)$$

Here $\hat{\sigma}_i$ and $\hat{t}_i$ are Pauli matrices which act in the Dirac band and valley (K and K') subspaces, respectively, $v_D$ is the phase velocity of the modes at the Dirac points which is a photonic analogue of the Dirac velocity, and $\zeta \sim \chi_{xy}$ characterizes the strength of the coupling between TE and TM polarizations induced by the bi-anisotropy. The Hamiltonian given in Eq. (1) operates in the space spanned by the 8-component wavefunctions $V = [V_K, V_{K'}]$ where $V_K = [E_z^I(K), E_z^{II}(K), H_z^I(K), H_z^{II}(K)]^T$, and the superscripts *I* and *II* are the bands' subspace indices. From the structure of the Hamiltonian one can see that the observed gap opening is due to the coupling of (*i*) the first TE and the second TM Dirac bands and (*ii*) the first TM and the second TE Dirac bands, respectively. By applying an additional transformation $\widehat{U} = [\hat{I}, \hat{I}; \hat{I}, -\hat{I}]$, where $\hat{I}$ is $2 \times 2$ identity matrix, the effective Hamiltonian in Eq. (1) can be block diagonalized

$$\widehat{\mathcal{H}} = \widehat{U}^+ \widehat{H} \widehat{U} = \begin{bmatrix} v_D(\hat{t}_z \hat{\sigma}_x k_x + \hat{\sigma}_y k_y) + \zeta \hat{t}_z \hat{\sigma}_z & 0 \\ 0 & v_D(\hat{t}_z \hat{\sigma}_x k_x + \hat{\sigma}_y k_y) - \zeta \hat{t}_z \hat{\sigma}_z \end{bmatrix} (2).$$

The new Hamiltonian $\widehat{\mathcal{H}}$ acts on the wavefunctions $\Psi = [\Psi_K, \Psi_{K'}]$ where $\Psi_K = [E_z^I(K) + H_z^I(K), E_z^{II}(K) + H_z^{II}(K), E_z^I(K) - H_z^I(K), E_z^{II}(K) - H_z^{II}(K)]^T$ and is identical to that introduced in Ref. [2] for graphene with spin-orbit coupling. Through this transformation we have obtained two decoupled photonic subsystems with four-component spinor wavefunctions of the form $\boldsymbol{\psi}^+ = [\boldsymbol{E}_z(K) + \boldsymbol{H}_z(K), \boldsymbol{E}_z(K') + \boldsymbol{H}_z(K')]$ and $\boldsymbol{\psi}^- = [\boldsymbol{E}_z(K) - \boldsymbol{H}_z(K), \boldsymbol{E}_z(K') - \boldsymbol{H}_z(K')]$, where bold symbols stand for 2-vectors in Dirac band subspace. These wavefunctions have one peculiar property; because the electric and magnetic field components of the electromagnetic wave transform differently under time-reversal ($\widehat{T} E_z = E_z$, while $\widehat{T} H_z = -H_z$), they are Kramer's partners of each other, $\widehat{T} \boldsymbol{\psi}^{+(-)} = \boldsymbol{\psi}^{-(+)}$. Components of these wavefunctions play the roles of the components of a photonic four-spinor and will be referred to as spin-up ($\boldsymbol{\psi}^+$) and spin-down ($\boldsymbol{\psi}^-$) photonic states. In the low-energy effective theory Eq. (2), the spin-up and spin-down sectors are described by the massive Dirac Hamiltonian with opposite signs of mass terms $\pm \zeta \hat{t}_z \hat{\sigma}_z$ characterizing opposite coupling to bi-anisotropy. This Hamiltonian is known to yield a quantized Hall response, as first pointed out in the seminal work by Haldane [42]. The Hall response can be readily understood as being caused by an effective magnetic field due to Berry curvature in momentum space that is generated by a nonzero mass. In our photonic system Eq. (2), the two photonic spin sectors experience effective magnetic fields that are equal in magnitude but opposite in sign. Electromagnetic wave propagation therefore reflects a quantized photonic spin Hall effect that is exactly analogous to the quantum spin Hall effect in electronic systems [2, 41].

We emphasize that our proposal is different from earlier ones. The pioneering studies of Ref. [13, 14, 16] proposed photonic crystal structures that break time-reversal symmetry and realize a quantum Hall response with one-way edge states. In contrast, our photonic meta-crystal structure preserves time-reversal symmetry and realizes quantum spin Hall response with spin-polarized one-way photonic edge transport. In the following, we demonstrate this by first calculating the bulk topological invariant and then calculating the propagating midgap modes along discontinuity-lines in the bi-anisotropy landscape.

To confirm the nontrivial topological character of the photonic states in the proposed bi-anisotropic meta-crystal, we have numerically calculated the Chern number [6] for each photonic band using the formula $C_n^\pm = \frac{1}{2\pi}\int_{BZ} d^2\mathbf{k}\left[\partial_{k_x}A_y^{n\pm}(\mathbf{k}) - \partial_{k_y}A_x^{n\pm}(\mathbf{k})\right]$, where $\mathbf{A}_n^\pm = -i\langle\boldsymbol{\psi}^{n\pm}(\mathbf{k})|\nabla_\mathbf{k}|\boldsymbol{\psi}^{n\pm}(\mathbf{k})\rangle$ is the Berry connection, $\boldsymbol{\psi}^{n\pm}(\mathbf{k})$ is the spin-up (+) and spin-down (−) eigenfunction for band $n$ obtained from the plane-wave expansion method, and the subscript BZ implies integration over the Brillouin zone. We find that as long as the bi-anisotropy of the form $\chi_{xy} = -\chi_{yx}$ is present, the doubly degenerate Dirac bands (corresponding to ± wavefunctions playing the role of Kramer's partners) acquire a finite Chern number of opposite signs $C_n^\pm = \pm 1$ (provided by equal $\pm 1/2$ contributions from the K and K′ points). Thus, while the net Chern number for a pair of degenerate bands vanishes since time-reversal symmetry is preserved, the system exhibits a topologically nontrivial phase which is the photonic analogue of the quantum spin Hall effect [43], characterized by a quantized spin Chern number [44] $C_{\text{spin}} = (C^+ - C^-) = 2$ and a $Z_2$ invariant [1, 5] of 1.

## 3. Edge states

To study edge-state transport in photonic meta-crystal structures, we consider a domain wall formed between two photonic band insulators that are topologically distinct. In electronic transport, interfacing a topological insulator with vacuum suffices to form a domain wall because vacuum is a good insulator for electrons. In photonic transport, on the other hand, vacuum is a good conductor for photons; therefore a photonic domain wall needs to be constructed between two photonic crystals with a full energy gap in the range of operating frequencies. We first study the case of a domain wall formed between two domains with opposite signs of magneto-electric coupling $\chi_{xy} > 0$ and $\chi_{xy} < 0$. We performed first-principle finite element method (FEM) numerical calculations for a super-cell of 30x1 unit cells of our original photonic meta-crystal structure with periodic boundary conditions imposed in $x$- and $y$-directions. The domain wall is then introduced by flipping the sign of $\chi_{xy}$ in the center of the super-cell. The resulting band diagram clearly shows a pair of edge states (Fig. 1c) localized near the domain wall (Fig. 1d). In close proximity to the K-point, the pair of spin-up ($\boldsymbol{\psi}_e^+$) and spin-down ($\boldsymbol{\psi}_e^-$) edge states are found to respect time-reversal symmetry with the same energy $\omega^+(\delta k_\|) = \omega^-(-\delta k_\|)$ and the same field amplitudes $|\boldsymbol{\psi}_e^+(\delta\mathbf{k})| = |\boldsymbol{\psi}_e^-(-\delta\mathbf{k})|$ (where $\delta\mathbf{k} = \mathbf{k} - \mathbf{k}_K$ denotes departure from the K-point in the reciprocal space). Remarkably, examination of the phase dynamics of the fields reveals that the two edge modes have clockwise and counter-clockwise rotating field patterns. Thus the photonic spin degree of freedom defined according to the eigenstates of the Hamiltonian Eq. (2) carries a clear physical meaning – it corresponds to the sense of self-rotation of the field $\boldsymbol{\psi}_e^\pm$ in real space. This confirms that the edge states are indeed time-reversed partners that satisfy $\hat{T}\boldsymbol{\psi}_e^{+(-)} = \boldsymbol{\psi}_e^{-(+)}$, self-rotate in opposite senses, and propogate along the edge in opposite directions. (See Multimedia Supplement: Animation).

Perhaps the most striking consequence of the topologically nontrivial phase is the presence of robust edge states at a boundary with trivial insulators. To confirm the existence of such states we consider an interface between the proposed photonic topological insulator and a topologically trivial photonic metamaterial possessing a complete photonic band gap in the part of spectrum enclosing the Dirac point (Supplement B). For the topologically trivial photonic region, we have used, for simplicity, the same set of parameters as in the nontrivial region, except with (i) a zero bi-anisotropy $\chi = 0$, and (ii) a smaller lattice constant (scaled down by 0.76) to ensure that the bulk-band Dirac point of the topologically nontrivial region lies within the energy range of the full band gap in the topologically trivial region. Scaling the lattice constant by a factor 0.76 matches a complete band gap spanning the frequency range $\Delta\omega \sim 0.3\text{-}0.33\ [2\pi c/a_0]$ (Fig. 1a) between the lowest frequency singlet and Dirac doublet of the topologically trivial meta-crystal to the bi-anisotropy induced band gap of the topologically nontrivial structure (Supplement B). To illustrate more vividly the presence of one-way spin-polarized edge states, we solve a harmonic propagation problem using a finite element method and selectively excite spin-up and spin-down edge modes in the spatially extended 2D structure. As Fig. 2a illustrates, when the electric and magnetic fields of the source that is placed at the center of the boundary between the topologically trivial and nontrivial photonic meta-crystals are in-phase ($E_z = H_z$) or out-of-phase ($E_z = -H_z$), the source selectively excites either spin-up (right propagating) or spin-down (left propagating) surface waves, respectively. This is consistent with our analysis from the low-energy effective model Eq. (2) and confirms that the interface supports two edge states connected by the time-reversal operation. Moreover, it was found that these modes exist regardless of the configuration of the interface for any separations and for both commensurate and incommensurate junctions of the crystals. Thus the modes exhibit a high degree of tolerance to local configurations of the interface, confirming their topological origin.

To confirm robustness of the edge states with respect to structural imperfections, we have modeled several types of defects. The first type represents a sharp bend of the interface between the trivial and nontrivial crystals formed by stacking the two structures. As one can see from the top subplot in Fig. 2b, the forward propagating spin-up edge state $\psi_e^+$ is perfectly transmitted by both sharp corners of the zigzag-shaped interface regardless of the different stacking configurations at every segment along the zigzag. The field configuration conforms to every segment and no reflection into the backward-propagating (spin-down $\psi_e^-$) edge mode takes place. The second type of defect investigated is a cavity formed by the absence of columns in both crystals in the vicinity of the interface (Fig. 2c). In this case we find that the edge mode strongly interacts with the cavity but is eventually re-radiated into the same direction as a forward-propagating spin-up mode without any reflection. We observe the same behavior for arbitrary cavity size with different number of columns removed at different locations. Finally, perhaps the most intriguing type of disorder we have studied is a strong local distortion of the crystalline lattice introduced by randomly shifting rods in both crystals in the proximity of their junction, modeling a local fusing of the two crystals. However, even in this case the edge mode demonstrates complete tolerance and avoids back-scattering by detouring around the disordered region. Similar simulations were performed for the backward-propagating spin-down edge mode for all three types of defects and exactly the same behavior, robustness and tolerance against the structural imperfections (bottom plots in Figs. 2b-d), was found.

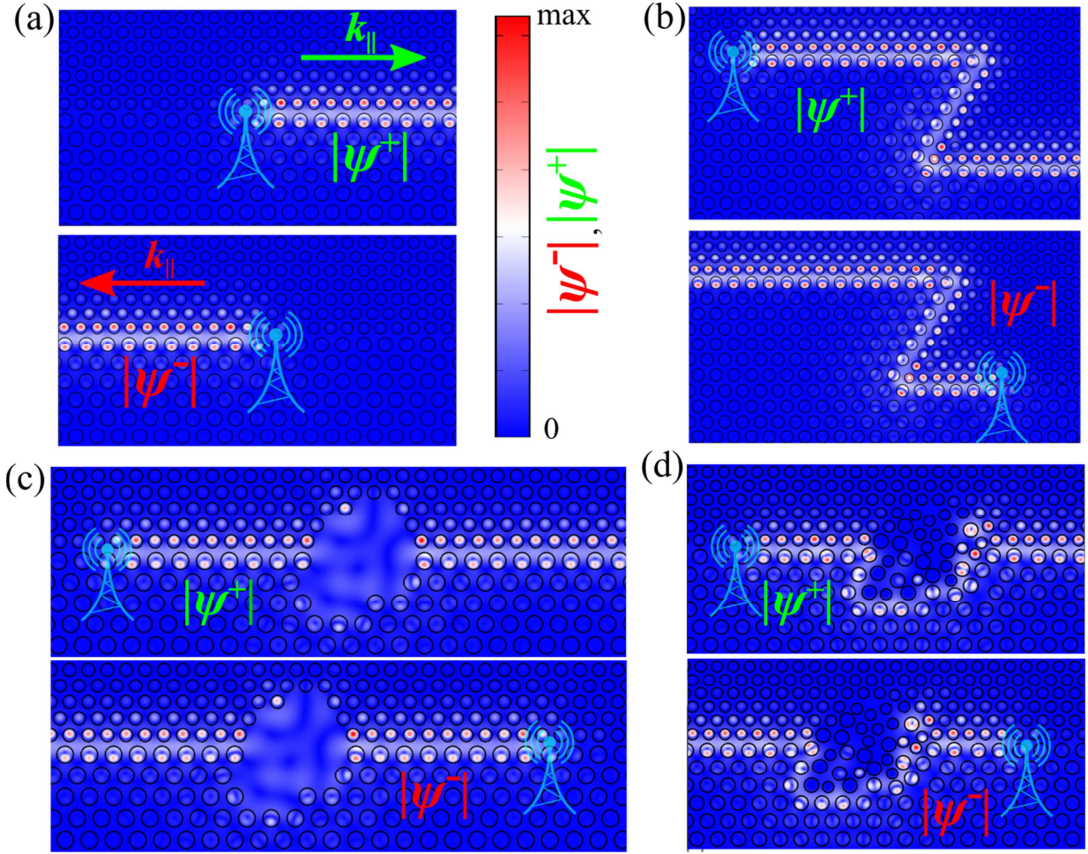

Figure 2: (a) Selective excitation of spin-up and spin-down photonic one-way edge states at the interface between topologically trivial and nontrivial photonic insulators. The color scale indicates the local field intensity $|\pmb{\psi}_e^\pm(x,y)|$ in plots (a)-(d). (b)-(d) Robustness of the edge modes against different types of defects: (b) sharp bending of the interface, (c) cavity obstacle, and (d) strongly disordered domain in both of the adjacent crystals.

## 4. Possible experimental implementation

To achieve the desirable $\epsilon$-$\mu$ matched bi-anisotropic metamaterial we propose here a design which is rather traditional for the microwave spectral range [45]. We consider a metamaterial composed of split-ring resonators with their axes in $x$-$y$ plane providing magnetic response $\mu_\perp \neq 1$, and resting on high permittivity slabs (Figure 3). This structure can be designed such that in the frequency range of interest $\mu_\perp \approx \epsilon_\perp \approx 14$. This ensures that we will have nearly degenerate Dirac points for TE and TM polarizations. We note that according to our simulations, a slight $\epsilon$-$\mu$ mismatch $\delta = \mu_\perp - \epsilon_\perp$ can be tolerated for sufficiently large $\chi_{xy}$, and robust edge states survive as long as the gap due to the bi-anisotropy exceeds the spectral detuning between the TE and TM bands caused by a finite value of $\delta$. This situation can be described by the effective Hamiltonian Eq. (2) with the introduction of an additional term of the form $\Delta \hat{H} \sim \delta\, \hat{\sigma}_0 \hat{s}_x$, where $\hat{\sigma}_0$ is an identity matrix acting on the band subspace and $\hat{s}_x$ is the spin subspace Pauli's matrix.

The presence of the split-ring resonators also implies a non-vanishing bi-anisotropy. In the structure shown in Fig. 3 having the same orientation of the split–rings lying on $z$-$x$ and $z$-$y$ planes ensures the required antisymmetry, $\chi_{xy} = -\chi_{yx}$. Indeed, as can be seen from Fig. 3(a), for the split ring lying on the $x$-$z$ plane, the magnetic field $H_y$ induces an electric moment $\boldsymbol{d}_x$ *collinear* to $x$-direction, while for the identical split-ring lying on the $y$-$z$ plane, the magnetic field $H_x$ induces an electric moment $\boldsymbol{d}_y$ *anti-collinear* to $y$-direction. Thus, by stacking many layers of such subwavelength split-ring resonators in both $x$ and $y$ directions one can simultaneously obtain both bi-anisotropy and magnetic response. Note, however, that in practice it might be better to separate sources of magnetism and bi-anisotropy since the gap width is proportional to $\zeta \sim \chi_{xy}/\epsilon_\perp \mu_\perp$ (Supplement A), and therefore high values of $\epsilon_\perp$ and $\mu_\perp$ diminish the effects of bi-anisotropy. Thus it is preferable to induce bi-anisotropy in the low-index background or have specially designed core-shell rods with the outer layer providing strong bi-anisotropy but relatively low electric and magnetic polarizabilities. Finally we mention that a topological phase similar to that reported here should also be observable in a wider class of bi-anisotropic systems with quadratic degeneracies [ 14, 46].

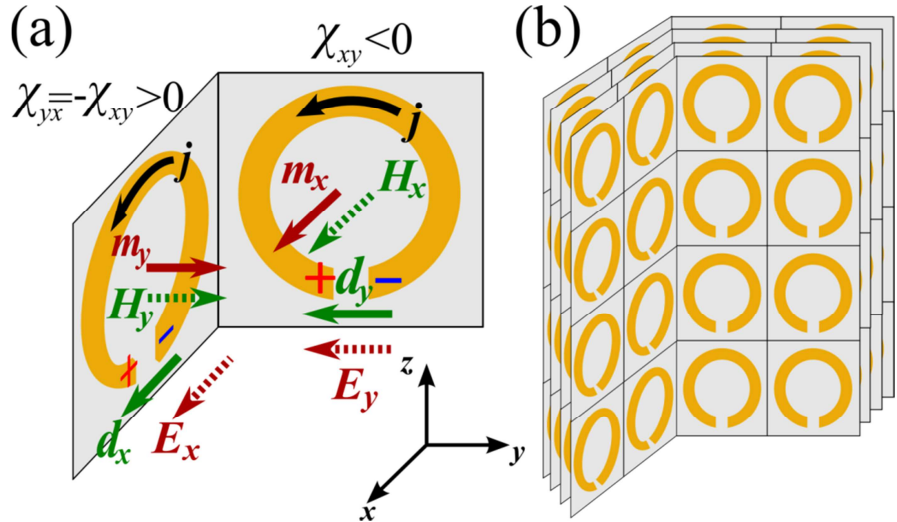

Figure 3: (a) Schematics illustrating the bi-anisotropic response of split-ring resonators lying on orthogonal (*zx* and *zy*) planes. (b) Illustration of a three-dimensional uniaxial magetic/bi-anisotropic metamaterial that can be used for making a photonic topological insulator.

## 5. Summary

In summary, we have shown that the photonic analogue of two-dimensional topological insulators can be realized in a photonic meta-crystal built from $\epsilon$-$\mu$ matched bi-anisotropic periodic metamaterial. We have demonstrated that the proposed photonic meta-crystal structure exhibits a topologically nontrivial photonic quantum spin Hall phase. Topologically protected spin-polarized one-way edge states were shown to exist at interfaces formed between metamaterials with reversed bi-anisotropy and at interfaces between topologically trivial and nontrivial structures. These states are reciprocal (bi-directional) and are related to each other by time-reversal symmetry and robust against different types of defects and disorder. Our proposed system is particularly attractive because it can be implemented without magneto-optical or gyromagnetic materials which are invariably lossy. Realizing robust topological states in photonic systems should open up new possibilities for engineering topological properties that are useful for applications, particularly where tolerance to disorder is essential.